\documentclass[english]{article}%
\usepackage[utf8]{inputenc}
\usepackage[T1]{fontenc}
\usepackage{graphicx}
\usepackage{authblk}
\usepackage{ulem}
\usepackage{amsmath,amssymb,amsfonts}
\usepackage{amsmath}
\usepackage{amsfonts}
\usepackage{amssymb}%
	\title{Extended quasilocal Thermodynamics of Schwarzchild-anti de Sitter
	black holes}
\date{}
	\author[1,*]{W. B. Fontana}
	\author[1,$\dagger$]{M. C. Baldiotti}
	\author[2]{R. Fresneda}
	\author[3]{C. Molina}
	\affil[1]{Departamento de F\'isica, Universidade Estadual de Londrina, 86051-990, Londrina-PR, Brazil.}
	\affil[2]{Centro de Matem\'atica, Computação e Cogniç\~ao, Universidade Federal do ABC, Av. dos Estados 5001, 09210-580, Santo Andr\'-SP, Brazil.}
	\affil[3]{Escola de Artes, Ci\^encias e Humanidades, Universidade de S\~ao Paulo, Av. Arlindo Bettio 1000, CEP 03828-000, São Paulo-SP, Brazil.
		
	\textit{Electronic Adress: $^{*}$weslei@uel.br; $^{\dagger}$baldiotti@uel.br; $^{2}$rodrigo.fresneda@ufabc.edu.br; $^{3}$cmolina@usp.br}}

\begin{document}


\maketitle

	
	
\begin{abstract}
In this work we study a homogeneous and quasilocal Thermodynamics
associated to the Schwarzschild-anti de Sitter black hole. The usual
thermodynamic description is extended within a Hamiltonian approach
with the introduction of the cosmological constant in the thermodynamic
phase space. The macroscopic treatment presented is consistent in as much as it
respects the laws of black hole Thermodynamics and accepts the introduction
of any thermodynamic potential. We are able to construct new equations
of state that characterize the Thermodynamics. Novel phenomena can
be expected from the proposed setup.
\end{abstract}

\section{Introduction}

Over the last few decades, research on black hole Physics has suggested
fundamental connections between Gravity, Thermodynamics and Quantum Field
Theory \cite{Padmanabhan-2010}. Theoretical evidences indicate that black hole
Thermodynamics is a key to this relationship, dictating the laws which black
holes obey and, at the same time, showing that those laws can be seen as a
thermodynamic description of gravitational systems
\cite{Padmanabhan-2010,Jacobson,Padmanabhan2015}.

Since the pioneering work of Bekenstein \cite{key-1}, it is well-known that
black holes possess an entropy proportional to their horizon area. Afterwards
Hawking, Carter and Bardeen provided a general proof of the laws of black hole
Mechanics \cite{3}. A genuine black hole Thermodynamics was derived from
Hawking's result of black hole evaporation \cite{2}, which showed that black
holes emit radiation due to semiclassical effects, and thus have an associated
physical temperature. Since then, black hole Thermodynamics has become a major
topic in Physics, with special attention to the thermodynamic properties of
asymptotically anti de Sitter (AdS) geometries (see for example \cite{5,4})
due to the developments in anti de Sitter/conformal field theories (AdS/CFT)
correspondences \cite{Madacena,Witten,Gubser}.

In the present work we focus on the thermodynamic behavior of \linebreak%
$D$-dimensional Schwarzchild-anti de Sitter black holes (SAdS). We consider a
macroscopic setup which arises from a description where the SAdS black hole is
in equilibrium with a thermal atmosphere that is generated by Hawking emission
\cite{5}. We will refer to minimal SAdS Thermodynamics as the description
where there is only one thermodynamic variable, the entropy (or the horizon
radius). In such a scenario, there is no distinction between isothermic and
isentropic processes, and the notion of a thermodynamic temperature is not
well-defined \cite{8}. Besides, in this minimal setup, the first law of
Thermodynamics is not consistent with the Smarr formula \cite{6}. Therefore,
the minimal description is not a thermodynamic theory in the standard sense.
These issues can be fixed once we extend the thermodynamic theory in order to
have additional degrees of freedom. So, in the case of a SAdS black hole, this
amounts to a Thermodynamics which is closer to that of usual matter systems
\cite{7}.

The extension of the SAdS black hole Thermodynamics, obtained by introducing
the cosmological constant as a dynamical variable, furnishes a good path
towards a consistent theory. In this formalism $\Lambda$ is interpreted as a
pressure and the black hole mass as the enthalpy, instead of the internal
energy \cite{12}. The cosmological constant is needed as a thermodynamic
parameter {for the theory to have homogeneous equations of state. Homogeneity
is required for extensivity \cite{6} and for the existence of an integrating
factor for the reversible heat exchange \cite{14}. However, homogeneity
requires that $\Lambda$ must be introduced in the theory in a very specific
manner, otherwise inconsistencies appear (for example, in the construction of
the thermodynamic potentials, as pointed out in \cite{15}). In particular, if
one tries to use the cosmological constant as a thermodynamic variable in this
extension, the obtained theory will present singularities in the Legendre
transformation between $\Lambda$ and its conjugate variable \cite{13}. }

Even after the construction of a consistent theory, there is a problem related
to the physical interpretation of the thermodynamic temperature. The usual
definition of black hole temperature is based on the horizon surface gravity.
But this quantity is not uniquely defined in stationary asymptotically anti de
Sitter geometries, since there is no preferred normalization for the
associated time-like Killing vector field.

An attempt to solve this ambiguity problem was proposed in \cite{16,17}. The
authors assumed a quasilocal approach, where physical quantities are defined
on a bounded spacetime region $R$. In this way the time-like Killing vector
field has no divergences and it is well-defined. This procedure defines new
quantities on the boundary, such as an energy function $E$ that plays the role
of the internal energy in the thermodynamic description. As one is dealing
with a bounded region, the quasilocal approach introduces a new thermodynamic
variable connected with the position $R$ of the observer in spacetime. As a
result, the temperature function depends on this new variable. In this
formulation, the temperature is redshifted to zero as $R\rightarrow\infty$ due
to the associated Tolman factor \cite{18}, which can be interpreted as a
manifestation of the confining character of the AdS asymptotics. Although the
quasilocal Thermodynamics gives a better definition of the temperature in a
non-asymptotic flat spacetime, it still lacks homogeneity, and all related
problems remain.

In a previous work \cite{15}, we extend the minimal setup using a Hamiltonian
approach to Thermodynamics \cite{19}, where the equations of state are
realized as constraints on a symplectic phase space (see also \cite{28} for an
approach based on the Hamilton-Jacobi formalism). We are able to incorporate
$\Lambda$ in a consistent manner in \cite{15}, solving the homogeneity issue
and obtaining a sensible Smarr formula for the SAdS Thermodynamics. However,
we are not able to fix the functional form of $\Lambda$ in \cite{15}, as it
remains dependent on free parameters of the model. In this work we construct
an extended version for the quasilocal thermodynamic description using the
aforementioned Hamiltonian method. The approach is to construct a macroscopic
effective description where semiclassical effects are implicitly assumed,
although no microscopic formalism (e.g. quantum field theories) is used. The
resulting theory has none of the issues described above and gives a
well-defined interpretation for the thermodynamic temperature. We show that in
this formalism it is possible to promote $\Lambda$ to a thermodynamic
variable. Besides, using geometric and thermodynamic arguments, we propose
some functional forms for the cosmological constant.

The structure of this work is as follows. In section \ref{sec:Minimal-AdS-} we
review the minimal description of the $D$-dimensional SAdS black hole. In
section \ref{sec: cosmolog const} we give a brief account of the extensions
which use the cosmological constant as a thermodynamic variable. In section
\ref{FB} we discuss the quasilocal approach to black hole Thermodynamics. In
section \ref{sec:Extended-AdS} the quasilocal thermodynamic description is
extended by means of the Hamiltonian approach. In the process, a new equation
of state for the black hole is constructed. Stability issues are studied in
section \ref{sec:Stability}. In section \ref{sec:Conclusions}, we discuss the
final considerations for this work and some paths that can be followed in the
future. Throughout this paper, we use the metric signature $\left(
-++\cdots+\right)  $ and natural units with $G=\hbar=c=k_{B}=1$.

\section{\label{sec:Minimal-AdS-} Minimal Schwarzchild-AdS Thermodynamics}

The Schwarzchild-anti de Sitter spacetime is the spherically symmetric vacuum
solution of the Einstein's equations with a negative cosmological constant
$\Lambda$. Its metric can be written as
\begin{equation}
ds^{2}=-N^{2}\left(  r\right)  dt^{2}+\frac{1}{N^{2}\left(  r\right)  }%
dr^{2}+r^{2}d\Omega_{D-2}^{2}\,, \label{eq:2.1.1}%
\end{equation}
where $d\Omega_{D-2}^{2}$ is the squared line element of the $(D-2)$%
-dimensional sphere $S^{D-2}$. The lapse function $N(r)$ in~(\ref{eq:2.1.1})
is expressed in terms of the (negative) cosmological constant $\Lambda$ and a
(positive) parameter $M$ as
\begin{equation}
N^{2}\left(  r\right)  =1-\frac{16\pi M}{\left(  D-2\right)  B_{D}}\frac
{1}{r^{D-3}}-\frac{2\Lambda r^{2}}{\left(  D-1\right)  \left(  D-2\right)
}\,, \label{eq:2.1.2}%
\end{equation}
where $B_{D}$ denotes the canonical volume of $S^{D-2}$, namely,
\begin{equation}
B_{D}=\frac{2\pi^{\frac{D-1}{2}}}{\Gamma\left(  \frac{D-1}{2}\right)  }\,.
\end{equation}
It is convenient to define an effective cosmological constant $\tilde{\Lambda
}$ that is related to $\Lambda$ as
\begin{equation}
\tilde{\Lambda}\equiv\frac{2\Lambda}{\left(  D-1\right)  \left(  D-2\right)
}\,. \label{lambAdsR}%
\end{equation}

Staticity of SAdS spacetime is characterized by the existence of a global
Killing vector field $\partial/\partial t$, which is time-like outside the
black hole. However, since the geometry is not asymptotically flat, there is
no preferred normalization for the Killing field. In fact, its norm
\begin{equation}
\left\vert g\left(  \frac{\partial}{\partial t},\frac{\partial}{\partial
t}\right)  \right\vert =N^{2}(r) \label{norm}%
\end{equation}
diverges as $r\rightarrow\infty$.

The Schwarzchild-anti de Sitter spacetime has a Killing horizon at $r=r_{+}$
associated to $\partial/\partial t$. The horizon radius $r_{+}$ is the unique
positive root of $N(r)$. A surface gravity $\kappa$ and a surface area $A$, or
more precisely a $(D-2)$-volume, can be associated to the horizon, whose
expressions in terms of $r_{+}$ are
\begin{equation}
\kappa=\frac{1}{2}\left[  \frac{D-3}{r_{+}}-\left(  D-1\right)  \tilde
{\Lambda}r_{+}\right]  \ ,~A=B_{D}r_{+}^{D-2}\,. \label{eq:2.1.5}%
\end{equation}
Using the expressions~(\ref{eq:2.1.5}), one obtains the following relation
among the parameters $M$, $A$ and $\kappa$,
\begin{equation}
8\pi\frac{D-1}{D-2}M=A\left(  \frac{A}{B_{D}}\right)  ^{\frac{1}{D-2}}+\kappa
A\,. \label{eq:2.9}%
\end{equation}

The existence of a Killing horizon implies a thermodynamic description for the
Schwarzchild-anti de Sitter spacetime \cite{27}. We will apply methods of
classical mechanics and thermodynamics on an effective theory, where it is
implicitly assumed that semiclassical effects generate a thermal Hawking
atmosphere \cite{5,15,20} around the black hole, and that it is in thermal
equilibrium with its atmosphere. In the minimal description, there is only one
thermodynamic variable: the entropy $S$. Hence, the fundamental equation has
the form $U=U\left(  S\right)  $, where $U$ denotes the system's internal
energy. Concretely, the minimal Schwarzchild-anti de Sitter Thermodynamics is
defined characterizing the internal energy $U$, the entropy $S$ and the
temperature $T$ as
\begin{equation}
U\equiv M,\quad S\equiv\frac{A}{4},\quad T=\frac{\kappa}{2\pi}\,.
\label{eq:2.1}%
\end{equation}
In fact, using the construction~(\ref{eq:2.1}), one verifies that
\begin{equation}
\frac{\partial U}{\partial S}=T\Rightarrow dU=TdS\,,
\end{equation}
which is the first law of Thermodynamics in the minimal scenario.

Relation~(\ref{eq:2.9}) can be rewritten in terms of the thermodynamic
variables $U$, $S$ and $T$ as
\begin{equation}
\frac{D-1}{D-2}U-TS=\frac{1}{2\pi}\left(  \frac{B_{D}}{4}S^{D-3}\right)
^{\frac{1}{D-2}}\,. \label{eq:2.10}%
\end{equation}
The above expression is the only equation of state in the minimal
thermodynamic description of SAdS black holes.

Despite its simplicity, minimal SAdS Thermodynamics has significant drawbacks.
For $\Lambda=0$ one can fix the normalization~(\ref{norm}) such that $g\left(
\partial/\partial t,\partial/\partial t\right)  =-1$ at the spatial infinity
and the obtained theory is homogeneous. While, for the SAdS spacetime, there
is no preferred normalization for the time-like Killing field. Thus, there is
no unique definition of temperature, as a Killing field proportional to
$\partial/\partial t$, will have a surface gravity different from $\kappa$ in
(\ref{eq:2.1.5}). Also, for $\Lambda\neq0$, the internal energy is no longer
homogeneous and the minimal Thermodynamics is not standard \cite{14,15}.

\section{\label{sec: cosmolog const}The cosmological constant as a
thermodynamic variable}

In this subsection we review the first attempts at generalizing the minimal
AdS Thermodynamics. The possibility of treating the cosmological constant as a
dynamical variable was proposed in \cite{11,BroTe88}, by coupling the
gravitational field to tensor fields. A quasilocal treatment was developed in
\cite{CreMa95}, by considering an Einstein-Hilbert-Dilaton action coupled to
various types of Abelian and non-Abelian gauge fields. These works do not
address the problem of the conjugate variable to $\Lambda$ in a thermodynamic
interpretation, neither do they consider different thermodynamic descriptions,
which can be obtained by Legendre transformations of $\Lambda$ and its
conjugated pair.

The internal energy $M$ in the minimal AdS Thermodynamics (as defined in
Section \ref{sec:Minimal-AdS-}) is not a first order homogeneous function. In
the asymptotically flat case, Euler's theorem for homogeneous functions
provides a route between the first law of black hole mechanics and the Smarr
formula \cite{24} for stationary black holes \cite{GauMyT99,TowZa2001}.
However, when $\Lambda\neq0$, in order to apply this procedure, one must take
into account the scaling properties of the cosmological constant
\cite{CalGoK2000}. By direct calculation for specific (A)dS black hole
solutions, expressions for the Smarr formula and the first law with a variable
$\Lambda$ have been obtained in \cite{CalGoK2000,9,10,WanWuXD2006,CarGr2008}.
The first law can also be obtained using Hamiltonian perturbation theory
techniques \cite{Tra85,TraFo20014,SudWa92}. Besides the scaling argument
\cite{6}, homogeneity of the equations of state is required for a consistent
black hole Thermodynamics \cite{14,23}.

Using the Killing potential introduced in \cite{BazZy90,Kas2008}, and the
techniques introduced in \cite{9}, it is possible to extend the Smarr formula
to the $D$-dimensional SAdS scenario. It follows that, for the case of the AdS
black hole with no charge nor angular momentum, one has the $D$-dimensional
Smarr formula \cite{kastor2010},
\begin{equation}
\left(  D-3\right)  M=\left(  D-2\right)  \frac{\kappa A}{8\pi}+2\frac
{\theta\Lambda}{8\pi}\,\,. \label{smarr}%
\end{equation}
In expression~(\ref{smarr}), $\theta$ represents a new thermodynamic variable,
conjugated to $\Lambda$. Using~(\ref{eq:2.1.5}), (\ref{eq:2.9}) and the Smarr
formula~(\ref{smarr}) for $D$ dimensions, one obtains
\begin{equation}
\theta=\frac{B_{D}}{D-1}\left(  \frac{4S}{B_{D}}\right)  ^{\frac{D-1}{D-2}%
}=\frac{B_{D}}{D-1}r_{+}^{D-1}\,. \label{vol}%
\end{equation}

Result~(\ref{vol}) identifies $\theta$ with the \textquotedblleft volume
extracted from spacetime\textquotedblright\ by the black hole \cite{6}. This
identification suggests that the cosmological constant $\Lambda$ should be
interpreted as a pressure and the black hole mass $M$ should be identified
with its enthalpy $H$ \cite{6}. In this context, there are two independent
thermodynamic variables, such that
\begin{equation}
H\equiv M\,,\,\,S\equiv\frac{A}{4}\,,\,\,T\equiv\frac{\kappa}{2\pi}%
\,,\,\,\Pi\equiv-\frac{\Lambda}{8\pi}\,\,, \label{new_thermo}%
\end{equation}
where $\Pi$ is the thermodynamic pressure associated to $\theta$. A different
approach was proposed in \cite{AbbDe82}, where conserved charges in AdS were
analyzed. When the cosmological constant is considered as a thermodynamic
variable, new phenomena appear in the description of black holes. For example,
these objects behave like a Van der Waals fluid, with properties sometimes
encountered in more commonplace scenarios \cite{KubMa2012}. One often calls
black hole chemistry the setup where $\Lambda$ is a thermodynamic variable
\cite{KubMa2015,Ma2016} (see \cite{4} for a review).

Despite the intriguing arguments defining $M$ as the enthalpy, there is an
inconsistency in the procedure. The problem is due to the fact that $\theta$
does not depend on $\Pi$, which leads to a singularity in the Legendre
transformation of the pair $\left(  \Pi,\theta\right)  $. As a matter of fact,
from~(\ref{new_thermo}) one sees that
\begin{equation}
\left.  \frac{\partial H}{\partial\Pi}\right\vert _{S}=\theta\,.
\end{equation}
Considering now~(\ref{smarr}) and~(\ref{eq:2.9}), one determines the internal
energy $U$,
\begin{equation}
U=H-\Pi\theta=\left(  D-2\right)  \frac{\,B_{D-2}}{16\pi}\left(  \frac
{4S}{B_{D-2}}\right)  ^{\frac{D-3}{D-2}}\,, \label{u}%
\end{equation}
and it follows that~(\ref{u}) does not give the first law, that is,
\begin{equation}
\frac{\kappa}{2\pi}=T\neq\frac{\partial U}{\partial S}\,.
\end{equation}
We stress that this problem exists only when the black hole possesses no other
characteristic besides its mass. When charge or angular momentum are present,
the thermodynamic potential can be constructed by thermodynamic \cite{13} or
statistical mechanics arguments \cite{CalGoK2000}, and no singularity is
present. In the present work we will consider spherically symmetric and
electrically neutral spacetimes. For this class of black holes, we will
construct a singularity-free thermodynamic description.

\section{\label{FB}Quasilocal Schwarzchild-AdS Thermodynamics}

To solve some of the problems of the minimal Thermodynamics, one possible
approach is a quasilocal treatment, where the quantities of interest are
defined in a bounded region. More specifically, we will use the general
formalism presented by Brown and York \cite{17}, which was later applied to
asymptotic anti de Sitter geometries by Brown, Creighton and Mann \cite{16}.
It is worth mentioning other approaches based on extensions of the Noether
theorem, presented for instance in \cite{barnish} and references therein.

In the Brown-York approach, one considers a bounded region $\mathcal{M}%
\subset\mathcal{S}$ of a spacetime $\mathcal{S}$ with (pseudo-Riemannian)
metric $g_{\mu\nu}$ in $D$ dimensions. It is assumed that $\mathcal{M}$ has
the topology $\Sigma\times I$, where $\Sigma$ is a space-like $(D-1)$%
-dimensional hypersurface and $I$ is an interval of the real line. The region
$\mathcal{M}$ is composed by three pieces: ``initial'' and ``final''
space-like hypersurfaces $\mathcal{T}_{I}$ and $\mathcal{T}_{F}$; and a
hypersurface $\mathcal{B}$ which connects $\mathcal{T}_{I}$ and $\mathcal{T}%
_{F}$. It follows that $\mathcal{B}=B\times I$, where $B$ is a compact
$(D-2)$-dimensional space-like surface. That is, $\mathcal{B}$ is foliated
into $(D-2)$-surfaces. The induced metric on $B$ is denoted by $\sigma_{ab}$.
Neither global hyperbolicity nor specific asymptotic structures are assumed
for $\mathcal{S}$.

Brown and York employed a Hamilton-Jacobi analysis of the action functional
describing Einstein's gravity in spacetimes with finite boundaries. The
appropriate gravitational action for this description is \cite{21}
\begin{equation}
S_{g}=S_{0}+\frac{1}{\alpha}S_{L}\,, \label{eq:3.1}%
\end{equation}
with $\alpha$ being a coupling constant and the action $S_{L}$ being given by
\begin{align}
S_{L}=\int_{\mathcal{M}}d^{D}x\sqrt{-g}\left(  \frac{\mathcal{R}}{2}%
-\Lambda\right)  +\int_{\mathcal{T}_{F}}d^{D-1}x\sqrt{h^{F}}K^{F}\nonumber\\
-\int_{\mathcal{T}_{I}}d^{D-1}x\sqrt{h^{I}}K^{I}-\int_{\mathcal{B}}%
d^{D-1}\sqrt{-\gamma}\Theta\,. \label{action_SL}%
\end{align}
In~(\ref{eq:3.1}), $S_{0}$ is a functional of the metric on $\partial
\mathcal{M}$ such that $\delta S_{0}$ vanishes when the metric is fixed on the
boundary $\partial\mathcal{M}$. The induced metrics on $\mathcal{T}_{I}$,
$\mathcal{T}_{F}$ and $\mathcal{B}$ are denoted by $h_{ij}^{I}$, $h_{ij}^{F}$
and $\gamma_{ij}$ respectively. The curvature scalar on $\mathcal{M}$ is
denoted by $\mathcal{R}$. The quantities $K^{I}$, $K^{F}$ and $\Theta$ are the
traces of the extrinsic curvatures in the hypersurfaces $\mathcal{T}_{I}$,
$\mathcal{T}_{F}$ and $\mathcal{B}$ respectively.

Decomposing the metric $\gamma_{ij}$ in the ADM form \cite{22}, one obtains
that
\begin{equation}
\gamma_{ij}dx^{i}dx^{j}=-N^{2}dt^{2}+\sigma_{ab}\left(  dx^{a}+V^{a}dt\right)
\left(  dx^{b}+V^{b}dt\right)  \,, \label{eq:3.5}%
\end{equation}
where $N$ is the lapse function and $V^{a}$ is the shift vector. In the
Brown-York approach, the main contribution from the variation of $S_{g}$ comes
from the boundary $\mathcal{B}$. Denoting $u_{i}$ as the unit normal vector to
$B$ and $\sigma_{i}^{a}=\delta_{i}^{a}$ the tensor that projects covariant
tensors from $\mathcal{B}$ to $B$, it follows that
\begin{equation}
\delta S_{g}|_{\mathcal{B}}=\int_{\mathcal{B}}d^{D-1}x\sqrt{\sigma}\left[
j_{a}\delta V^{a}-\varepsilon\delta N+\left(  \frac{N}{2}\right)  s^{ab}%
\delta\sigma_{ab}\right]  \,, \label{eq:3.7}%
\end{equation}
where
\begin{align}
\varepsilon &  =\frac{1}{\alpha}k-\varepsilon_{0}\,,~j_{i}=-\frac{2}{\sqrt{h}%
}\sigma_{ij}P^{jk}n_{k}-\left(  j_{0}\right)  _{i}\,,\label{epsilon}\\
s^{ab}  &  =\frac{1}{\alpha}\left(  k^{ab}+\left(  n_{\mu}a^{\mu}-k\right)
\sigma^{ab}\right)  -\left(  s_{0}\right)  ^{ab}\,. \label{eq:3.13}%
\end{align}
In expressions~(\ref{eq:3.7})-(\ref{eq:3.13}), $n_{\mu}$ represents the
components of the unit normal to $\mathcal{B}$, $k^{ab}$ is the extrinsic
curvature of the boundary $B$ and $k$ its trace, $P^{ij}$ denotes the
gravitational momentum for the hypersurface $\Sigma$ and $a^{\mu}=u^{\nu
}\nabla_{\nu}u^{\mu}$ indicating the acceleration of $u^{\mu}$. The terms with
index zero in~(\ref{epsilon}) and~(\ref{eq:3.13}) are proportional to the
functional derivatives of $S_{0}$.

From~(\ref{eq:3.7}), it follows that the term $-\sqrt{\sigma}\varepsilon$ is
equal to the time rate of change of the action, wherein the lapse function
controls the changes of $\mathcal{B}$ with respect to time. Thus,
$\varepsilon$ is identified as an energy surface density and the total
quasilocal energy is defined by integration over the $(D-2)$-dimensional
surface $B$ \cite{17},
\begin{equation}
E=\int_{B}d^{D-2}x\sqrt{\sigma}\varepsilon\,. \label{eq:3.14}%
\end{equation}

Let us analyze the $D$-dimensional Schwarzchild-anti de Sitter black hole. In
$D$ dimensions, we set the coupling constant to $\alpha=8\pi$. Since the
geometry is spherically symmetric, it is natural to consider the surfaces $B$
as $(D-2)$-spheres, specified by $r=R$, with constant $R$. Staticity of the
SAdS geometry allows the identification of $\mathcal{T}_{I}$ and
$\mathcal{T}_{F}$ with the hypersurfaces $t=t_{I}$ and $t=t_{F}$. Given the
choices made, the energy density $\varepsilon$ associated to the surface $r=R$
is
\begin{equation}
\varepsilon=-\frac{D-2}{8\pi R}\sqrt{1-\frac{16\pi M}{\left(  D-2\right)
B_{D}R^{D-3}}-\tilde{\Lambda}R^{2}}-\varepsilon_{0}\left(  R\right)  \,.
\label{eq:3.2.2}%
\end{equation}
Integrating over $B$, one obtains the total energy:
\begin{equation}
E=-R^{D-2}B_{D}\left[  \frac{D-2}{8\pi R}\sqrt{1-\frac{16\pi M}{\left(
D-2\right)  B_{D}R^{D-3}}-\tilde{\Lambda}R^{2}}+\varepsilon_{0}\left(
R\right)  \right]  \,. \label{eq:3.2.3}%
\end{equation}
We note that some quantities, such as $E$ in~(\ref{eq:3.2.3}), depend on the
arbitrary function $\varepsilon_{0}$.

Brown, Creighton and Mann \cite{16} assumed that the quantity $E$ in
(\ref{eq:3.2.3}) represents the internal energy of the thermodynamic system.
Following \cite{16}, the surface pressure $P$ at the boundary $r=R$ can be
obtained for the $D$-dimensional Schwarzschild-anti de Sitter spacetime:
\begin{align}
P  &  \equiv-\frac{\partial E}{\partial\left(  B_{D}R^{D-2}\right)
}\nonumber\\
&  =\frac{R\left(  D-3\right)  }{8\pi N(R)}\left[  R^{-2}-\frac{8\pi MR^{1-D}%
}{\left(  D-2\right)  B_{D}}-\frac{D-2}{D-3}\tilde{\Lambda}\right]
+\frac{\partial\left(  R^{D-2}\varepsilon_{0}\right)  }{\partial\left(
R^{D-2}\right)  }\,. \label{eq:pressure}%
\end{align}
In the same fashion, a local temperature $T_{L}$ can be found as
\begin{equation}
T_{L}\equiv\frac{\partial E}{\partial S}=\frac{D-3}{4\pi N(R)}\left[  \left(
\frac{4S}{B_{D}}\right)  ^{-\frac{1}{D-2}}-\frac{D-1}{D-3}\tilde{\Lambda
}\left(  \frac{4S}{B_{D}}\right)  ^{\frac{1}{D-2}}\right]  \,.
\label{eq:3.2.4}%
\end{equation}

The local temperature $T_{L}$ in~(\ref{eq:3.2.4}) is damped by a term
proportional to the inverse of the lapse function
\begin{equation}
T_{L}=\frac{1}{N\left(  R\right)  }\frac{\kappa}{2\pi} = \frac{T}{N\left(
R\right)  } \,.
\end{equation}
The damping of a local temperature by the Tolman factor, as observed by Brown,
Creighton and Mann in their approach, is a generic characteristic of any
reasonable Thermodynamics in a static spacetime \cite{18}.

Thus, in the quasilocal description of Schwarzschild-anti de Sitter spacetime,
the temperature measured by an observer at $r=R$ is a well-defined quantity.
Also, the Thermodynamics is constructed with two degrees of freedom, since the
position $R$ is introduced as a new thermodynamic variable. Two of the
problematic issues in the minimal SAdS thermodynamic description are solved.

Still, the problem associated to homogeneity remains in the SAdS
Thermodynamics proposed by Brown, Creighton and Mann. Indeed, the lapse
function $N(r)$ must be a homogeneous function of order zero. This will be
true only if $M/R^{D-3}$ and $\Lambda R^{2}$ also are homogeneous functions of
order zero, as it is apparent from~(\ref{eq:2.1.2}). In order to recover
homogeneity, $\Lambda$ will be considered a thermodynamic parameter with the
help of the formalism developed in \cite{15,19}.

\section{\label{sec:Extended-AdS}Extended quasilocal SAdS Thermodynamics}

\subsection{Extended phase space}

The Hamiltonian approach to Thermodynamics consists of the identification of
thermodynamic variables with local coordinates $\left(  q,p\right)  $ of a
phase space. In this manifold, a Hamilton function $h\left(  q,p,t\right)  $
is introduced, and the symplectic structure is given by $\omega=d\theta$,
where $\theta=pdq$ is the tautological one-form. Equations of state are
realized as constraints and the one-form $\theta$ on the surface of
constraints becomes the differential of a thermodynamic potential. In this
way, one shows that Legendre transformations of thermodynamic potentials
amount to canonical transformations in phase space. Furthermore, one can
extend this description by the addition of new coordinates and momenta
$\left(  \tau,\xi\right)  $. This is done without increasing the number of
physical degrees of freedom, by supplying a Hamiltonian constraint
$H=\xi+h\left(  q,p,\tau\right)  $. As a result, the tautological one-form
$\theta$ becomes the Poincar\'{e}-Cartan form $pdq-hdt$ on the constraint
surface. For further details, see \cite{19}.

Let us first rename the thermodynamic variables as
\begin{equation}
q=\frac{4S}{B_{D}},\quad p=\pi T_{L}=\frac{\kappa}{2N},\quad x=R^{D-2}%
,\quad\varpi=-4\pi P\,. \label{4.5}%
\end{equation}
They label local coordinates on an open region of $\mathbb{R}^{4}$ with
symplectic structure given by the tautological one-form $\theta=pdq+\varpi
dx$. The equations of state~(\ref{eq:pressure}) and~(\ref{eq:3.2.4}) become constraint equations
$\phi_{1}=0$ and $\phi_{2}=0$, which are written in terms of the mechanical
variables~(\ref{4.5}) as
\begin{align}
\phi_{1}  &  =p-\frac{D-3}{4N}\left(  q^{-\frac{1}{D-2}}-\frac{D-1}%
{D-3}q^{\frac{1}{D-2}}\tilde{\Lambda}\right)  \,,\nonumber\\
\phi_{2}  &  =\frac{D-3}{4N}\left\{  2x^{-\frac{1}{D-2}}+x^{-1}\left[  \left(
q^{\frac{D-1}{D-2}}-2\frac{D-2}{D-3}x^{\frac{D-1}{D-2}}\right)  \tilde
{\Lambda}-q^{\frac{D-3}{D-2}}\right]  \right\} \nonumber\\
&  +\varpi+4\pi\frac{\partial\left(  x\varepsilon_{0}\right)  }{\partial x}\,.
\label{4.7}%
\end{align}
We will consider the energy~(\ref{eq:3.2.3}) a thermodynamic potential, so one
has a fundamental equation. Then, the form $\theta$ is the differential $dE$
on the constraint surface $\phi_{1}=\phi_{2}=0$, $\left.  \theta\right\vert
_{\phi_{1}=\phi_{2}=0}=dE$. The symplectic structure in the phase space
$\left(  q,\,p;x,\,\varpi\right)  $ is defined by the two-form $\omega
=d\theta$. According to the canonical Poisson structure given by $\omega$, the
Poisson brackets between the constraints vanish identically, so the set of
constraints is first-class. Therefore, there are no physical degrees of
freedom, as expected from the general theory \cite{19}. We shall extend the
description given in section \ref{FB} by introducing a new canonical pair
$\left(  \xi,\tau\right)  $ of mechanical variables, such that the
tautological one-form $\theta$ becomes
\begin{equation}
\theta\mapsto\tilde{\theta}=\frac{B_{D}}{4\pi}\left(  pdq+\varpi dx\right)
+\xi d\tau\,, \label{ext-tauto form}%
\end{equation}
with $\xi=\partial E / \partial\tau$ on the constraint surface. We follow
\cite{15} in that $\Lambda$ is considered a function on phase space, depending
on the coordinates $q$, $x$ and $\tau$. In particular, $\varepsilon_{0}$ and
$N$ depend on these coordinates through $\Lambda$. Since the tautological
one-form $\tilde{\theta}$ must reproduce the differential $dE$ on the
constraint surface, one is led to a new constraint,
\begin{align}
\phi_{3}  &  = \xi- \frac{\partial E}{\partial\tau}\nonumber\\
&  =\xi+\frac{B_{D}\left(  D-2\right)  }{16N\pi}\left(  q^{\frac{D-1}{D-2}%
}-x^{\frac{D-1}{D-2}}\right)  \frac{\partial\tilde{\Lambda}}{\partial\tau
}+xB_{D}\frac{\partial\varepsilon_{0}}{\partial\tau}\,. \label{phi3}%
\end{align}
With this, the notion of thermodynamic energy is not the same as the one in
the original theory. In addition, the constraints~(\ref{4.7}) receive
contributions coming from derivatives of $\Lambda$ and $\varepsilon_{0}$ with
respect to the coordinates, which in terms of thermodynamic quantities are
described by
\begin{align}
&  P\mapsto P+\frac{D-2}{16\pi N}\left[  \left(  \frac{4S}{B_{D}}\right)
^{\frac{D-1}{D-2}}-R^{D-1}\right]  \frac{\partial\tilde{\Lambda}}%
{\partial\left(  R^{D-2}\right)  }\,,\label{ext-pressure}\\
&  T_{L}\mapsto T_{L}-B_{D}\frac{D-2}{16\pi N}\left[  \left(  \frac{4S}{B_{D}%
}\right)  ^{\frac{D-1}{D-2}}-R^{D-1}\right]  \frac{\partial\tilde{\Lambda}%
}{\partial S}-B_{D}R^{D-2}\frac{\partial\varepsilon_{0}}{\partial
\tilde{\Lambda}}\frac{\partial\tilde{\Lambda}}{\partial S}\,.
\label{ext-temperature}%
\end{align}
The total set of constraints is again first-class, resulting in a theory with
no physical degrees of freedom. We note that if the cosmological constant
depends on the entropy, the temperature $T_{L}$ no longer has the usual
geometric interpretation as surface gravity.

Now we impose homogeneity to the equations of state, which is a necessary
condition in order to have a consistent black hole Thermodynamics
\cite{14,23}. Although homogeneity does not completely fix the functional form
of $\Lambda$, it does impose strong conditions on it. Considering the
extensivity of the entropy \linebreak$S\rightarrow\lambda S$
in~(\ref{eq:3.2.4}), one has
\begin{equation}
T_{L}\rightarrow\lambda^{-\frac{1}{D-2}}T_{L}\,,
\end{equation}
and, since the lapse function~(\ref{eq:2.1.2}) is homogeneous of order zero,
we obtain
\begin{align}
R  &  \rightarrow\lambda^{\frac{1}{D-2}}R,\quad E\rightarrow\lambda
^{\frac{D-3}{D-2}}E,\quad P\rightarrow\lambda^{-\frac{1}{D-2}}P\,,\nonumber\\
\tilde{\Lambda}  &  \rightarrow\lambda^{-\frac{2}{D-2}}\tilde{\Lambda}%
,\quad\xi\rightarrow\lambda^{\frac{D-3}{D-2}-c}\xi,\quad\tau\rightarrow
\lambda^{c}\tau\,. \label{homogeneity}%
\end{align}
Once the lapse function $N$ does not depends explicitly on $\tau$, but only
through $\Lambda\left(  S,R,\tau\right)  $, it is not possible to fix the
homogeneity order of $\tau$ by demanding that $N$ be homogeneous of order
zero. So, $c$ remains an arbitrary real constant. This constant connects all
the thermodynamic descriptions that respect the homogeneity condition.

Consider the Euler theorem for homogeneous functions \cite{6}. If a function
scales as $G\left(  \lambda^{\alpha_{1}}x_{1},\,\lambda^{\alpha_{2}}%
x_{2},\,\lambda^{\alpha_{3}}x_{3}\right)  =\lambda^{\eta}G$, then
\begin{equation}
\eta G=\alpha_{1}\frac{\partial G}{\partial x_{1}}x_{1}+\alpha_{2}%
\frac{\partial G}{\partial x_{2}}x_{2}+\alpha_{3}\frac{\partial G}{\partial
x_{3}}x_{3}\,. \label{eq:4.2.1}%
\end{equation}
Renaming $G=\tilde{\Lambda}$, $x_{1}=\tau$, $x_{2}=S$, $x_{3}=R^{D-2}$, one
has $\eta=2/\left(  2-D\right)  $, $\alpha_{1}=c$ and $\alpha_{2}=\alpha
_{3}=1$. With this we obtain
\begin{equation}
\tilde{\Lambda}=\frac{2-D}{2}\left(  c\frac{\partial\tilde{\Lambda}}%
{\partial\tau}\tau+\frac{\partial\tilde{\Lambda}}{\partial S}S+\frac
{\partial\tilde{\Lambda}}{\partial R^{D-2}}R^{D-2}\right)  \,.
\label{eq:4.2.2}%
\end{equation}
If $c=0$, there is no phase transition \cite{15}. This is unexpected,
considering Hawking-Page theory, and so we discard this case.

The general solution for $c\neq0$ of the differential equation~(\ref{eq:4.2.2}%
) is
\begin{equation}
\tilde{\Lambda}=\tau^{-\frac{2}{D-2}\frac{1}{c}}\mathcal{F}\left(  z,y\right)
\,,\quad z=S\tau^{-\frac{1}{c}}\,,\quad y=R^{D-2}\tau^{-\frac{1}{c}}\,,
\label{eq:4.2.3}%
\end{equation}
where $\mathcal{F}$ is any homogeneous function of order zero, as is clear
from the definition of its arguments above. Equation~(\ref{eq:4.2.3}) fixes
the functional form of the cosmological constant in the thermodynamic
description, solely on the grounds of the homogeneity condition.

We also observe that homogeneity relations~(\ref{homogeneity}) impose
limitations on the possible choices of $\varepsilon_{0}$, restricting it to
specific combinations of the thermodynamic quantities. These combinations need
to be homogeneous of order $1/(2-D)$ and, by the Euler theorem, $\varepsilon
_{0}$ must has the form
\begin{equation}
\varepsilon_{0}=\frac{1}{R}\mathcal{G}\left(  z\,,t\right)  \,,\quad
z=R\tau^{\frac{1}{c}\frac{1}{\left(  2-D\right)  }},\quad t=RS^{\frac{1}{2-D}%
}\,, \label{epsilon0}%
\end{equation}
where $\mathcal{G}$ is some arbitrary homogeneous function of order $0$.
Concerning the dependence of $\varepsilon_{0}$ on $z$ and $t$, it should be
taken into account that the cosmological constant is a function on phase
space, and so $\varepsilon_{0}$ depends at this point on $\tau$ and $S$
implicitly through $\Lambda$ (besides the explicit dependence on $R$). Another
result is that one can obtain the Smarr formula \cite{6,9,24} by using Euler's
formula~(\ref{eq:4.2.1}). Setting $G=E$, $x_{1}=S$, $x_{2}=R^{D-2}$,
$x_{3}=\tau$, we have
\begin{equation}
E=\frac{D-2}{D-3}\left(  T_{L}S-B_{D}PR^{D-2}+c\tau\xi\right)  \,,\quad
\xi=\frac{\partial E}{\partial\tau}\,. \label{eq:4.13}%
\end{equation}
Thus, the constant $c$ can be understood as a measure of the contribution to
the energy that arises due to the extension of phase space by the canonical
pair $(\tau,\xi)$. The $R$-dependence in~(\ref{eq:4.13}) implies that the
energy is dependent on the position of the observer at the boundary. In the
same way, if the AdS \textquotedblleft box\textquotedblright\ increases or
decreases in size (i.e., a change in the AdS radius $l=1/\sqrt{-\tilde
{\Lambda}}$), it will change the spacetime energy by an amount weighted by the
constant $c$.

\subsection{Fixing the Thermodynamics}

The function $\mathcal{F}$ in~(\ref{eq:4.2.3}) can be determined by demanding
that the cosmological constant is fixed by geometric arguments. Let us
consider the relation that determines the horizon radius, $N(r_{+})=0$, or
more explicitly,
\begin{equation}
1-\frac{16\pi M}{\left(  D-2\right)  B_{D}}\,\frac{1}{r_{+}^{D-3}}%
-\frac{2\Lambda}{\left(  D-1\right)  \left(  D-2\right)  }r_{+}^{2}=0\,.
\end{equation}
There are three parameters in this equation, and hence $r_{+}$ and $\Lambda$
can be chosen arbitrarily, as long as $r_{+}$ is positive and $\Lambda$ is
negative. Any choice of $r_{+}>0$ and $\Lambda<0$ generates a
Schwarzschild-AdS geometry. Since the entropy $S=B_{D}r_{+}^{D-2}/4$ is a
function of $r_{+}$ alone, any pair $(S,\Lambda)$ with $S>0$ and $\Lambda<0$
is associated with a SAdS spacetime, with $S$ and $\Lambda$ independent. Also,
a particular choice of a surface $r=R$ for the definition of the quasilocal
quantities should not depend on the background geometry (characterized by
$r_{+}$ and $\Lambda$, or by $S$ and $\Lambda$). Therefore, geometric
arguments suggest that thermodynamic variable $\Lambda$ should be independent
of $S$ and $R$. From (\ref{eq:4.2.3}) we see that the only way to satisfy
these requirements is if $\mathcal{F}=-K$ is constant, so
\begin{equation}
\tilde{\Lambda}=-K\tau^{-\frac{2}{D-2}\frac{1}{c}}\,.\label{l1}%
\end{equation}

Taking~(\ref{l1}) into account, $dE=\tilde{\theta}|_{{\{\phi\}}=0}$, where
$\{\phi\}$ is the total set of constraints, is expressed as
\begin{align}
dE  &  =\frac{B_{D}}{4\pi}\left(  pdq+\varpi dx\right) \nonumber\\
&  -\left[  \frac{B_{D}\left(  D-2\right)  }{16N\pi}\left(  q^{\frac{D-1}%
{D-2}}-x^{\frac{D-1}{D-2}}\right)  \frac{\partial\tilde{\Lambda}}{\partial
\tau}+xB_{D}\frac{\partial\varepsilon_{0}}{\partial\tau}\right]  d\tau\,,
\label{eq:4.10}%
\end{align}
and the additional constraint~(\ref{phi3}) becomes
\begin{equation}
\phi_{3}=\xi+\frac{B_{D}K\tau^{-\frac{2}{D-2}\frac{1}{c}-1}}{8Nc\pi}\left(
q^{\frac{D-1}{D-2}}-x^{\frac{D-1}{D-2}}\right)  +xB_{D}\frac{\partial
\varepsilon_{0}}{\partial\tau}\,. \label{eq:4.11}%
\end{equation}
One observes that the first law of Thermodynamics is given by
eq.~(\ref{eq:4.10}) evaluated at the constraint surface $\phi_{1}=\phi
_{2}=\phi_{3}=0$. The equation $\phi_{3}=0$ becomes, in thermodynamic
variables,
\begin{equation}
\xi=-\frac{2}{D-2}\frac{1}{Nc\tau}\frac{\Lambda}{8\pi}\left[  \theta\left(
R\right)  -\theta\left(  r_{+}\right)  \right]  -B_{D}R^{D-2}\frac
{\partial\varepsilon_{0}}{\partial\tau}\,, \label{xi-1}%
\end{equation}
where
\begin{equation}
\theta\left(  r\right)  =\frac{B_{D}}{D-1}r^{D-1}%
\end{equation}
is the Euclidian volume of a sphere of radius $r$ in $D$ dimensions. Since
$\Lambda/8\pi$ can be interpreted as the (volumetric) energy density of the
AdS space, it follows that, for $\varepsilon_{0}=0$, $cN\xi\tau$ is
proportional to the energy of the spacetime inside the radius $R$ sphere
discounted the volume $\theta(r_{+})$ that, as discussed in section
\ref{sec: cosmolog const}, can be identified as the volume extracted from
spacetime due to the presence of the black hole. That is the only quantity
with physical meaning. By substituting~(\ref{xi-1}) in (\ref{eq:4.13}), we obtain a
Smarr formula in terms of $R$, $S$ and $\Lambda$,
\begin{align}
\frac{D-3}{D-2}E  &  =T_{L}S-\frac{B_{D}\tilde{\Lambda}}{8\pi N}\left[
R^{D-1}-\left(  \frac{4S}{B_{D}}\right)  ^{\frac{D-1}{D-2}}\right]
-B_{D}R^{D-2}\left(  P+c\tau\frac{\partial\varepsilon_{0}}{\partial\tau}\right)
\nonumber\\
&  =T_{L}S-\frac{2\Lambda}{8\pi}\frac{\theta\left(  R\right)  -\theta\left(
r_{+}\right)  }{\left(  D-2\right)  N}-B_{D}R^{D-2}\left(  P+c\tau
\frac{\partial\varepsilon_{0}}{\partial\tau}\right)  \,.
\label{eq-E-lambdatil}%
\end{align}
The second expression in eq.~\eqref{eq-E-lambdatil} can be compared with the
analogous result in \cite{6}. The choice of a specific value of $c$ is a
matter of convenience. As shown in \cite{15}, it is not possible to use
$\Lambda$ as a thermodynamic variable in the simple extension of the minimal
SAdS Thermodynamics. This is because the conjugate variable $\xi$ will not
depend on $\tau$ and the Legendre transformation between the pair (necessary
to construct the internal energy from the enthalpy) is singular \cite{13,26}.
However, in the quasilocal approach discussed here, it is clear
from~(\ref{phi3}) that, even if $\tau=\Lambda$, $\xi$ still depends on $\tau$
(or $\Lambda$) through the lapse function $N\left(  S,R,\tau\right)  $ in the
denominator of the first term in the right hand side of~(\ref{phi3}), even
though the derivative $\partial\Lambda/\partial\tau$ is constant. So, in the
present setup, it is possible to promote $\Lambda$ to a thermodynamic variable
without introducing singularities. This choice corresponds to
\begin{equation}
c=-\frac{2}{D-2}\,. \label{c}%
\end{equation}

By setting $K=16\pi/\left(  D-1\right)  \left(  D-2\right)  $ in~(\ref{l1}),
we have that $\tau=\left\vert \Lambda\right\vert /8\pi$ is the energy density
of spacetime, as in Dolan's original proposal \cite{12,13}. Besides, for
$\varepsilon_{0}=0$, the conjugate variable $\xi=\theta\left(  R\right)
-\theta\left(  r_{+}\right)  $ is the total geometric volume of the local
description, discounted the black hole volume (taking into account the
correction by the lapse function). Or, in other words, $\xi$ is the total
geometric volume measured by an observer characterized by $N=1$. We stress
that, with the treatment presented here, it is possible to promote $\Lambda$
to a thermodynamic variable, as was originally proposed in \cite{9,10,11,13}.

Although interesting, there is nothing fundamental in the choice of $c$
in~(\ref{c}) from the thermodynamic point of view. A better strategy might be
to use the arbitrariness in $c$ in order to simplify a specific problem. For
example, one can choose to work with the fundamental equation $U\left(
S,R,\xi\right)  =E-\xi\tau$,
\begin{equation}
U=\left(  D-2\right)  \left\{  \frac{B_{D}}{8\pi N}\left[  \left(  \frac
{4S}{B_{D}}\right)  ^{\frac{D-3}{D-2}}-R^{D-3}\right]  -\left[  c+\frac
{1}{D-2}\right]  \tau\xi\right\}  \,.
\end{equation}
In this case, the expression can be simplified by setting
\begin{equation}
c=-\frac{1}{D-2}\,. \label{c2}%
\end{equation}
That is, using $\sqrt{-\Lambda}$ as a thermodynamic variable instead of
$\Lambda$. With this choice
\begin{equation}
U=\left(  D-2\right)  \frac{B_{D}}{8\pi N}\left[  \left(  \frac{4S}{B_{D}%
}\right)  ^{\frac{D-3}{D-2}}-R^{D-3}\right]  \,.
\end{equation}

The choice~(\ref{c2}) has the additional advantage that, as in the usual
thermodynamics, all the variables can be classified in two types: the
extensive variables $\left(  S,B_{D}R^{D-2},\xi\right)  $, of degree one, and
the intensive variables $\left(  T,P,\tau\right)  $, of degree $1/\left(
2-D\right)  $. We note that, in this case, $\xi$ is an extensive variable, so
$E$ is a function of two extensive and one intensive variable, and could be
interpreted as the enthalpy (as proposed in \cite{12,13} for the extension of
the minimal setup). In addition, $U$ depends only on the extensive variables
and can be identified with the internal energy (a quantity ill-defined in
\cite{12,13}). Other choices of $c$ could prove to be more convenient for
different thermodynamic potentials.

Some more comments about the choice of $c$ can be made. The first law of
Thermodynamics (\ref{eq:4.10}) alone cannot fix the parameter $c$, since it
only fixes the homogeneity of the product $\xi d\tau$. And according to the
homogeneity relations eq.~(\ref{homogeneity}), this product will be
independent of $c$. In fact, the only physical quantity independent of $c$ is
the product $\tau\xi$. The physical meaning of variables $\left(  \tau
,\,\xi\right)  $ can only be seen after the fixation of $c$. One example was
discussed in this section, eq.~\eqref{c2}, with $\tau=1/l$. Another example is
the case $c=(D-1)/(D-2)$, where $\tau$ can be seen as the effective volume in
the anti de Sitter space \cite{Elias}.

It should be noted that the asymptotic behavior of the theory depends heavily
on the choice of the arbitrary function $\varepsilon_{0}$, as it was pointed
out in \cite{16}. In our treatment, the only restriction upon the
$\varepsilon_{0}$ function is that it must obey eq.~(\ref{epsilon0}). The
asymptotic behavior of $P$ in eq.~(\ref{eq:pressure}) and $\xi$ in
eq.~(\ref{xi-1}) changes with different choices of $\varepsilon_{0}$, but once
we fix it there is no ambiguity.

Besides completely defining the Thermodynamics in terms of the parameters
$\Lambda,R$ and $r_{+}$, the above results guarantee the homogeneity condition
and a singularity-free Thermodynamics. As a consequence, any thermodynamic
potential can be used in this description (e.g., the enthalpy, the Helmholtz
free energy and so on), since generic Legendre transformations linking these
potentials are possible, as in Thermodynamics of non-gravitational systems.

\section{\label{sec:Stability}Stability conditions}

In this section we analyze the stability conditions for the system. We have a
consistent thermodynamic description for the SAdS black hole, characterized by
the equations of state for $P$ and $T_{L}$ in~(\ref{eq:pressure}) and
(\ref{eq:3.2.4}) respectively. In addition, there is the constraint
(\ref{eq:4.11}):
\begin{equation}
\xi=\frac{B_{D}K\left(  D-2\right)  }{16\pi N}\left[  \left(  \frac{4S}{B_{D}%
}\right)  ^{\frac{D-1}{D-2}}-R^{D-1}\right]  -R^{D-2}B_{D}\frac{\partial
\varepsilon_{0}}{\partial\tau}\,. \label{eq:4.3.3}%
\end{equation}
In~(\ref{eq:4.3.3}) we considered the fixed value~(\ref{c}) for the constant
$c$. We stress that there is nothing special with this choice and one could
work with the general expression~(\ref{xi-1}).

Let us consider the heat capacity
\begin{equation}
C_{R,\Lambda}=\left(  \frac{\partial E}{\partial T_{L}}\right)  _{R,\Lambda
}=T_{L}\left(  \frac{\partial S}{\partial T_{L}}\right)  _{R,\Lambda}\,,
\label{eq:4.3.4}%
\end{equation}
since our system has three degrees of freedom, and we consider processes with
two variables fixed. From~(\ref{eq:3.2.4}), it follows that
\begin{equation}
C_{R,\Lambda}=-\left(  D-2\right)  \frac{S}{\upsilon}\left[  D-3-\tilde
{\Lambda}\left(  D-1\right)  \left(  \frac{4S}{B_{D}}\right)  ^{\frac{2}{D-2}%
}\right]  \,, \label{eq:4.3.5}%
\end{equation}
where
\begin{equation}
\upsilon\left(  S,R,\Lambda\right)  =D-3+\tilde{\Lambda}\left(  D-1\right)
\left(  \frac{4S}{B_{D}}\right)  ^{\frac{2}{D-2}}-\frac{8\pi^{2}T_{L}^{2}%
}{R^{D-3}}\left(  \frac{4S}{B_{D}}\right)  ^{\frac{D-1}{D-2}}\,.
\label{eq:4.3.6}%
\end{equation}
The limit $R\rightarrow\infty$ of the function $\upsilon\left(  S,R,\Lambda
\right)  $ is
\begin{equation}
\lim_{R\rightarrow\infty}\upsilon=D-3+\tilde{\Lambda}\left(  D-1\right)
\left(  \frac{4S}{B_{D}}\right)  ^{\frac{2}{D-2}}\,. \label{eq:4.3.7}%
\end{equation}
Therefore, as $R$ grows, the heat capacity diverges as $\upsilon\rightarrow0$,
when
\begin{equation}
\tilde{\Lambda}\frac{D-3}{D-1}=-\left(  \frac{4S}{B_{D}}\right)  ^{\frac
{2}{D-2}}\,. \label{eq:4.3.8-1}%
\end{equation}
In terms of the horizon radius $r_{+}$ and the the cosmological constant
$\Lambda$, the previous equation assumes the form
\begin{equation}
r_{+}=r_{0},\quad r_{0}\equiv\sqrt{\frac{(D-3)(D-2)}{2|\Lambda|}}\,.
\label{eq:4.3.8}%
\end{equation}
We define a temperature $T_{SL}$ which is the quasilocal temperature
(\ref{eq:3.2.4}) taken at the critical radius~(\ref{eq:4.3.8}), which
coincides with the minimum of the temperature $T_{L}$ in the large $R$ limit.
Thus, if the temperature at the boundary is set to a value smaller than
$T_{SL}$, no black hole can exist, and the system will be dominated by thermal
radiation in an anti de Sitter background. On the other hand, if the
temperature is set to a value greater than $T_{SL}$, there will be two
solutions. Namely, a larger and stable black hole, with $r_{+}>r_{0}$ and
positive heat capacity; and a smaller and unstable black hole, with
$r_{+}<r_{0}$ and negative heat capacity. The temperature $T_{SL}$ is the
transition temperature for the Small-Large black hole transition \cite{25}.

Therefore, SAdS black holes can be in thermal equilibrium within an
environment with a negative cosmological constant when $R\rightarrow\infty$.
This was originally shown by Hawking and Page \cite{25} using a Euclidean path
integral approach. It is important to note that the Small-Large (SL) black
hole phase transition is obtained from the minimum of the Hawking temperature
(\ref{eq:2.1}) which, in general, is different from the minimum of the local
temperature $T_{L}$. The analysis we have made here is possible only in the
large $R$ limit, when the two minima asymptotically coincide and, thus, one
can read the SL phase transition from the temperature $T_{L}$.

We also note that, for the $D=3$ BTZ black hole, $\upsilon$ is independent of
$R$ and a straightforward calculation shows that $C_{R,\Lambda}=S$, which is
always positive. Thus, our description for the BTZ black hole has no local
instability, in agreement with \cite{26}.

\section{\label{sec:Conclusions}Conclusions and perspectives}

The first attempts at describing the thermodynamic behavior of SAdS black
holes were not entirely consistent \cite{15}. The minimal description lacks
homogeneity and the temperature is not well-defined, since the surface gravity
is not uniquely defined. The latter issue can be treated using quasilocal
quantities defined for $\left(  D-2\right)  $-spheres \cite{16,17,21}. With
this, one obtains a theory in which the temperature of the black hole depends
explicitly on the position of the observer.

However, the quasilocal formalism does not fix the homogeneity issue. We solve
this problem by using a Hamiltonian approach to Thermodynamics presented in
\cite{19}. We are able to find a new thermodynamic theory specifying the
entropy $S$, the observer position $R$ and the cosmological constant $\Lambda
$. We are also able to give an explicit form for the Smarr formula in terms of
these quantities. This result shows how the geometry contributes to the
internal energy of the thermodynamic description. It has all the properties of
the previously descriptions (i.e., has a well-defined temperature) and
furthermore is homogeneous. As a result, unlike all previous descriptions, the
presented development allows the use of any thermodynamic potential, which can
be obtained via Legendre transformations. Our approach brings the
Thermodynamics of Schwarzchild-anti de Sitter black holes closer to the
description of standard thermodynamic systems.

Besides predicting Small-Large black hole phase transitions, new effects are
expected, due to the existence of heat capacities associated to different
kinds of processes, as well as other equations of state which give analogs of
the compressibility coefficients of the standard theory. A natural development
of the present work would be the characterization of phase transitions and
critical exponents. Investigation along those lines is currently under way.

\section*{Acknowledgements}

W. B. F. acknowledges the support of the Coordination for the Improvement of
Higher Education Personnel (CAPES), Brazil with grants \#1591255 and
\#1775078. C. M. is supported by grant \#2015/24380-2, S\~{a}o Paulo Research
Foundation (FAPESP), Brazil; and grants \#307709/2015-9 and \#420878/2016-5,
National Council for Scientific and Technological Development (CNPq), Brazil.
R.F. acknowledges the support of S\~{a}o Paulo Research Foundation (FAPESP),
Brazil, with grant \#2016/03319-6.

\end{document}